\documentclass[reprint, aps, two column, showkeys]{revtex4-1}
\usepackage{graphicx,epsfig}
\usepackage{amssymb}
\usepackage{amsmath}
\usepackage{bm}
\usepackage{textcomp}
\usepackage[utf8]{inputenc}
\usepackage{color}

\begin{document}
\setcounter{page}{1}

\title[]{Cyclotron Orbits of Composite Fermions in the Fractional Quantum Hall Regime}
\author{Insun \surname{Jo}}
\author{Hao \surname{Deng}}
\author{Yang \surname{Liu}}
\author{L. N. \surname{Pfeiffer}}
\author{K. W. \surname{West}}
\author{K. W. \surname{Baldwin}}
\author{M. \surname{Shayegan}}
\affiliation{Department of Electrical Engineering, Princeton University, Princeton, NJ 08544, USA  }
\date{\today}

\begin{abstract}

We study a bilayer GaAs hole system that hosts two distinct many-body phases at low temperatures and high perpendicular magnetic fields. The higher-density (top) layer develops a Fermi sea of composite fermions (CFs) in its half-filled lowest Landau level, while the lower-density (bottom) layer forms a Wigner crystal (WC) as its filling becomes very small. Owing to the inter-layer interaction, the CFs in the top-layer feel the periodic Coulomb potential of the WC in the bottom-layer. We measure the magnetoresistance of the top-layer while changing the bottom-layer density. As the WC layer density increases, the resistance peaks separating the adjacent fractional quantum Hall states in the top-layer change nonmonotonically and attain maximum values when the cyclotron orbit of the CFs encloses one WC lattice point. These features disappear at \textit{T} = 275 mK when the WC melts. The observation of such geometric resonance features is unprecedented and surprising as it implies that the CFs retain a well-defined cyclotron orbit and Fermi wavevector even deep in the fractional quantum Hall regime, far from half-filling.

\end{abstract}

\maketitle 
When a strong perpendicular magnetic field (\textit{B}) quenches the kinetic energy of clean two-dimensional (2D) electrons at low temperature and the Coulomb interaction becomes dominant, the system exhibits a variety of quantum ground states at different filling factors, $\nu$ \cite{Tsui.PRL.1982,Perspective.Book,Shayegan.Review.2006,Jain.Book}. At very large \textit{B}, when $\nu < 1/5$, the electrons form a Wigner crystal (WC) \cite{Wigner.PR.1934,Lozovik.JETP.1975,Lam.PRB.1984,Levesque.PRB.1984,Andrei.PRL.1988,Jiang.PRL.1990,Goldman.PRL.1990,Li.PRL.1991,Li.PRL.1997,Li.PRB.2000,Shayegan.Review.1997,Chen.NatPhy.2006,Tiemann.NatPhy.2014,Deng.PRL.2016}, which manifests as an insulating phase in transport measurements because the WC is pinned to the ubiquitous disorder potential. At higher, odd-denominator fillings, the electrons condense into incompressible fractional quantum Hall states (FQHSs) \cite{Tsui.PRL.1982,Perspective.Book,Shayegan.Review.2006,Jain.Book}. These are elegantly described by composite fermions (CFs), quasi-particles consisting of an even number of flux quanta and an electron \cite{Jain.PRL.1989,Halperin.PRB.1993,Jain.Book}. Thanks to the flux attachment, at $\nu=1/2$ the CFs feel a zero effective magnetic field ($B^*$) and occupy a Fermi sea with a well-defined Fermi wavevector. This is evinced in experiments on samples with a periodic potential modulation where the geometric resonance (GR) of the CFs’ cyclotron orbit diameter with the period of the modulation is observed \cite{Jain.Book,Willett.PRL.1993, Kang.PRL.1993, Goldman.PRL.1994,Smet.PRL.1998, Kamburov.PRL.2012, Kamburov.PRB.2014, Kamburov.PRL.2014}. Further away from $\nu=1/2$, when $B^*$ becomes large, the CF picture accounts for the FQHSs as the integer QHS of CFs \cite{Jain.Book,Jain.PRL.1989}.

\begin{figure}[b!]
  \begin{center}
    \psfig{file=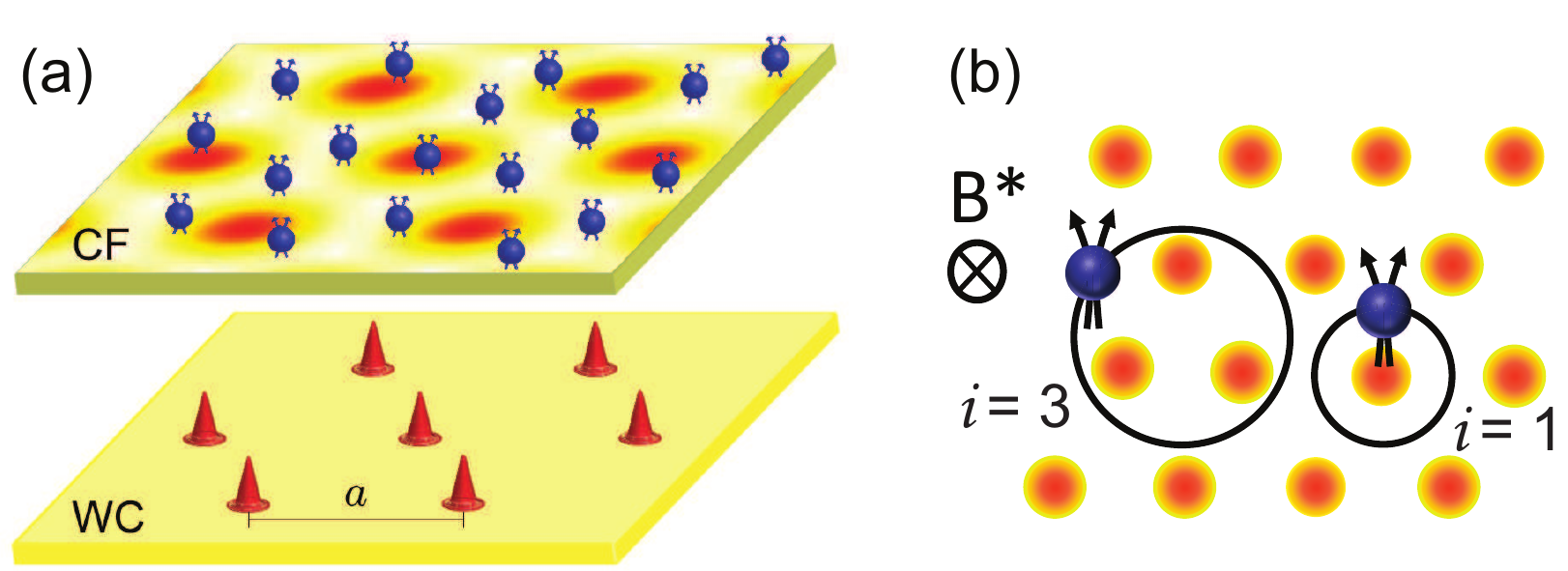, width=0.45\textwidth }
  \end{center}
  \caption{\label{fig1} (a) Schematic of a bilayer system consisting of CF (near $\nu=1/2$) and WC ($\nu \ll 1$) layers in a strong \textit{B}. The WC layer imposes a periodic potential modulation on the CF layer. (b) The CFs in a finite \textit{effective} magnetic field ($B^*$) exhibit a GR feature, i.e. a maximum in the magnetoresistance, when their cyclotron orbit encircles 1, 3, …  lattice points. In our sample we observe the $i=1$ GR.} 
\end{figure}
 
\begin{figure*}[t!]  
  \begin{center} 
    \psfig{file=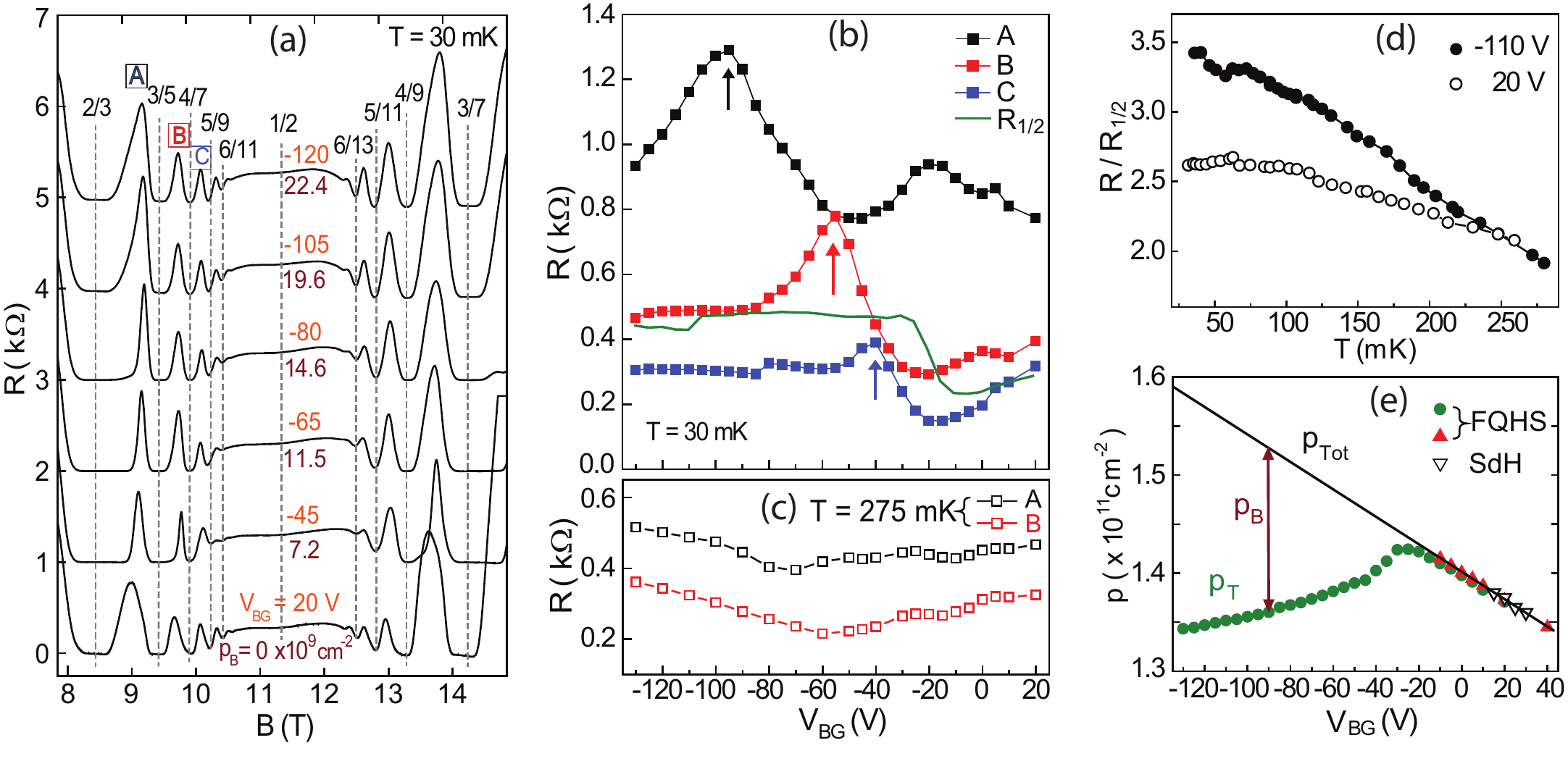, width=0.97\textwidth }
  \end{center}
  \caption{\label{fig2} (a) Magnetoresistance traces between $\nu=2/3$ and 3/7 at different $V_\mathrm{BG}$. The \textit{B}-axis scale is for the lowest trace (top-layer density $p_\mathrm{T}=1.42 \times 10^{11}$~cm$^{-2}$); \textit{B} for the other traces is normalized according to their measured $p_\mathrm{T}$ [see panel (e)] to align $\nu$. The other traces are shifted also vertically, and the applied $V_\mathrm{BG}$ and the measured bottom-layer density ($p_\mathrm{B}$) are indicated for each trace. (b) The evolution of the resistance peaks indicated as A, B, and C in (a) as well as the resistance at $\nu=1/2$ ($R_{1/2}$) are shown. The arrows indicate the expected positions satisfying the $i=1$ GR condition based on the measured $p_\mathrm{B}$ and $p_\mathrm{T}$. (c) The resistance of peaks A and B measured at 275 mK, illustrating that the $V_\mathrm{BG}$-dependent changes disappear. There is a similar disappearance for peak C. (d) Temperature dependence of peak A, normalized by $R_{1/2}$, is shown for $V_\mathrm{BG}=-110$ and 20 V. (e) The green circles and red triangles represent $p_\mathrm{T}$, deduced from the positions of the FQHSs at 30 and 275 mK, respectively; the empty triangles are $p_\mathrm{T}$ from the Shubnikov-de Haas oscillations measured at 30 mK. The total bilayer density $p_{Tot}$ is shown as the black line, and $p_\mathrm{B}$ is obtained by subtracting $p_\mathrm{T}$ from $p_\mathrm{Tot}$.}
\end{figure*}

A recent study of an asymmetric electron bilayer system hosting CFs in one layer and a WC in the other layer revealed a glimpse of the microscopic structure of the WC \cite{Deng.PRL.2016}. As schematically shown in Figs. 1(a) and 1(b), the Coulomb potential of the WC electrons imposes a periodic potential modulation on the adjacent CF layer. As a result, whenever the CFs’ cyclotron orbits enclose a certain number ($i=1,3,7$) of the potential modulation points, the CF layer's magnetoresistance exhibits a GR maximum. The $B^*$ positions of these maxima are consistent with the period of a WC arranged in a triangular lattice. Moreover, the GR features weaken with increasing temperature and disappear above $\sim$~200 mK as the WC melts. 

Here, we study a bilayer \textit{hole} system hosting a hole WC in one layer and hole-flux CFs in the other. Compared to the electron system, because of the larger effective mass and the stronger Landau level mixing for holes, the hole WC forms in the bottom-layer at much higher fillings \cite{ Shayegan.Review.1997,Santos.PRL.1992,Santos.PRB.1992,Csathy.PRL.2005}. The WC lattice at a higher filling means a smaller WC period, and thus causes the GR of CFs in the top layer to occur for smaller cyclotron orbits or, equivalently, for larger $B^*$. This leads to the GR features permeating into the regime where the strong FQHSs develop. Surprisingly, we observe that the resistance of the compressible states positioned between strong FQHSs changes with the WC layer density. In particular, it exhibits a maximum when the primary GR condition is satisfied, i.e., the CFs’ cyclotron orbit encloses one WC lattice point [see Fig. 1(b)]. The results have an intriguing implication: The CFs retain a well-defined Fermi wavevector even in the regime where strong FQHSs are observed.

The double-quantum-well structure used in our experiments is grown on a GaAs (001) substrate via molecular beam epitaxy. The two GaAs quantum wells are each 20 nm thick, and are separated by a 50-nm-thick Al$_{0.24}$Ga$_{0.76}$As barrier. The density of the top-layer ($p_\mathrm{T}$) is much larger than the bottom-layer density ($p_\mathrm{B}$) such that at a fixed, large magnetic field the top-layer is near its half-filling and hosts CFs while the bottom-layer has a filling $\ll 1$ and forms a WC [see Fig. 1(a)]. This bilayer structure with a large density imbalance is realized with a very asymmetric spacer and C $\delta$-layers; the thicknesses of the top and bottom Al$_{0.24}$Ga$_{0.76}$As spacer layers are 95 and 500 nm, respectively. The wafer is cut into a $4\times4$ mm$^2$ square shape, and In:Zn Ohmic contacts are made at the sample’s four corners. An In bottom-gate, covering the sample’s entire back, is used to apply a bias $V_\mathrm{BG}$ and tune $p_\mathrm{B}$.

Figure 2(a) shows the magnetoresistance traces measured at different $V_\mathrm{BG}$ at $T \simeq$~30 mK. The lowest trace is taken at $V_\mathrm{BG} = 20$~V, where the bottom-layer is completely depleted ($p_\mathrm{B}=0$) and $p_\mathrm{T}=1.42 \times 10^{11}$~cm$^{-2}$. It is similar to those for high-quality, single-layer 2D holes, and exhibits high-order FQHSs up to $\nu=6/13$. When a smaller $V_\mathrm{BG}$ is applied, holes start to occupy the bottom-layer. This leads to a weakening of the already fragile $\nu=5/9$, $6/11$, and $6/13$ FQHS minima at $V_\mathrm{BG} = -45$~V \cite{fnote1}. When $p_\mathrm{B}$ is further increased by lowering $V_\mathrm{BG}$, these FQHSs become strong again. Interestingly, the resistance peaks for the compressible states between $\nu=2/3,~3/5,~4/7$, and~$5/9$ FQHSs, marked as A, B, and C in Fig. 2(a), evolve non-monotonically as $V_\mathrm{BG}$ decreases. This is best seen in Fig. 2(b) where we show the measured resistance for peaks A, B, and C as a function of $V_\mathrm{BG}$. The resistance of peak A increases with decreasing $V_\mathrm{BG}$, reaches a maximum at $V_\mathrm{BG} \simeq -100$ V, and then it decreases \cite{fnote2}. Peaks B and C show a similar trend although they attain their maximum values at larger $V_\mathrm{BG} \simeq -55$ V and $-40$ V, respectively. We attribute the peaks observed in Fig. 2(b) to the GR of CFs in the top-layer, i.e., with their cyclotron orbit diameter matching the period of the WC formed in the bottom-layer [Fig. 1(b)]. The evolution observed in Fig. 2(b) comes about as the WC period that determines the condition for the CFs' GR changes with the WC density.
 
To support our conjecture, we first describe our precise determination of $p_\mathrm{B}$ and $p_\mathrm{T}$ using the mangetoresistance traces at various $V_\mathrm{BG}$. In Fig. 2(e), circles represent $p_\mathrm{T}$ determined from the field positions of the FQHS minima between $\nu=2/3$ and $3/7$. When only the top-layer is occupied $p_\mathrm{T}$ increases with decreasing $V_\mathrm{BG}$ linearly. For $V_\mathrm{BG}<-20$~V, as the bottom-layer starts to become occupied, $p_\mathrm{T}$ starts to \textit{decrease}. This negative compressibility, a manifestation of the strong exchange-correlation interaction, has been observed previously in various bilayer systems \cite{Eisenstein.PRB.1994,Ying.PRB.1995,Deng.PRL.2016}. The black solid line in Fig. 2(e) is a fit through the experimental data points for $V_\mathrm{BG}>-20$~V and represents the capacitive response of the total density of the bilayer ($p_\mathrm{Tot}$) when $V_\mathrm{BG}$ is applied; its slope is consistent with the expected geometrical capacitance of the sample. We obtain $p_\mathrm{B}$ by taking the difference between $p_\mathrm{Tot}$ and $p_\mathrm{T}$.

Based on $p_\mathrm{T}$ and $p_\mathrm{B}$ extracted from Fig. 2(e), we calculate the value of $V_\mathrm{BG}$ at which we expect the primary ($i=1$) GR resistance maximum to occur for states A, B, and C. As indicated by the arrows in Fig. 2(b), the measured resistance maxima for the compressible states A, B, and C indeed reach their peak values very close to where we expect the CFs to satisfy the GR condition, $2R^*=a$; here $a$ is the WC period, $R^*=\frac{\hbar k^*}{ e B^*}$ is the CFs’ cyclotron radius, and $k^*$ is their Fermi wavevector. We calculated $k^*$ assuming that the CFs are fully spin-polarized and their density is equal to $p_\mathrm{T}$, i.e., $k^*=\sqrt{4 \pi p_\mathrm{T}}$. Note also that for a triangular WC lattice, $a=\sqrt{\frac{2}{ \sqrt{3} ~p_\mathrm{B}}}$. The evolution seen in Fig. 2(b) is therefore quantitatively consistent with CF GR features evolving with $p_\mathrm{B}$ and $p_\mathrm{T}$. \textit{The GR features we observe near strong FQHSs imply that the CFs are present and appear to possess a well-defined Fermi wavevector, even at large values of $B^*$ where strong FQHSs are observed at adjacent fractional fillings.}

Before presenting more data, it is useful to discuss the role of disorder in our experiments. In Fig. 2(b) we also plot the resistance of our sample at $\nu=1/2$ ($R_{1/2}$). As $V_\mathrm{BG}$ is decreased from its highest value, $R_{1/2}$ initially decreases as $p_\mathrm{T}$ increases. Near the onset of the bottom-layer population (around $V_\mathrm{BG}$ = -15 V), $R_{1/2}$ suddenly starts to increase. This increase is likely related to the scattering of CFs off the holes in the bottom-layer which, at extremely low densities, is dominated by disorder. Such a phenomenon was indeed observed in bilayer systems at \textit{zero B}, and was attributed to enhanced electron scattering \cite{Gramila.PRL.1991, Katayama.PRB.1995}. For $V_\mathrm{BG} < -30$ V, however, $R_{1/2}$ is essentially constant \cite{fnote3}. This is in stark contrast to the non-monotonic behavior of the resistances of peaks A, B, and C in the same range of $V_\mathrm{BG} < -30$ V in Fig. 2(b). More importantly, the resistances of these peaks show clear maxima whose $V_\mathrm{BG}$ positions are consistent with the $i = 1$ GR of the CFs in the top-layer with the potential of a WC in the bottom-layer. We conclude that, while some level of disorder is present near the onset of the bottom-layer population, the peaks in Fig. 2(b) cannot be attributed to disorder.

Figure 2(c) illustrates the $V_\mathrm{BG}$ dependence of the resistance for states A and B measured at 275 mK. The maxima seen at $V_\mathrm{BG} \simeq -100$ V and -55 V at 30 mK [Fig. 2(b)] disappear at 275 mK, suggesting that the hole WC melts at this high \textit{T} and no longer induces a periodic potential modulation on the CF layer. Figure 2(d) depicts the \textit{T}-dependence of peak A's resistance at two fixed $V_\mathrm{BG}$. There is a much stronger \textit{T}-dependence for $V_\mathrm{BG}=-110$ V than for $V_\mathrm{BG}=20$ V. For the latter, the bottom-layer is completely depleted, and the system consists of only CFs in the top-layer. The decreasing resistance with increasing temperature can be attributed to the thermal broadening of the CF Landau levels as is well documented in single-layer systems \cite{Du.PRL.1993, Manoharan.PRL.1994}. However, at $V_\mathrm{BG}=-110$ V, where the height of peak A is influenced by the GR of CFs with the WC lattice [Fig. 2(b)], peak A’s resistance increases significantly with decreasing \textit{T} [Fig. 2(d)]. The increase is particularly notable below about 200 mK, which we associate with the melting temperature of the WC. This melting temperature is comparable to the one deduced from measurements in a CF-WC bilayer \textit{electron} system \cite{Deng.PRL.2016}; a quantitative comparison, however, is not warranted because of the different interlayer distance in the two systems, and also the much smaller $\nu$ in the electron WC case. 

 \begin{figure}[t!]
  \begin{center}
    \psfig{file=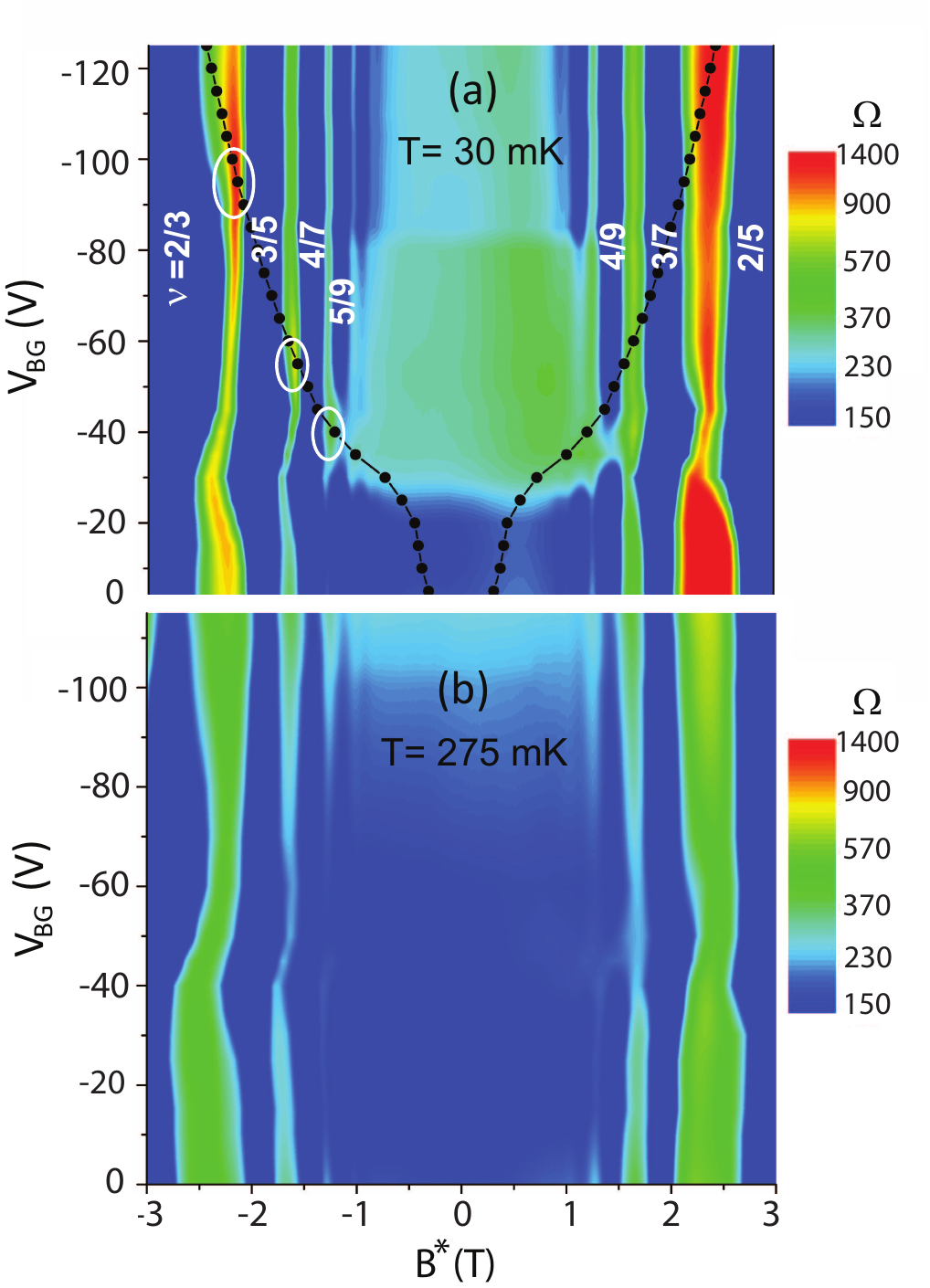, width=0.5\textwidth }
  \end{center}
  \caption{\label{fig3} (a) Summary of the magnetoresistance traces measured at 30 mK. The effective magnetic field $B^*$ for CFs is used for the horizontal axis. The black circles show the expected positions for the $i=1$ GR maximum. White ellipses mark the observed resistance maxima for peaks A, B, and C. (b) Results of similar measurements performed at 275 mK. }
\end{figure}

The plots in Fig. 3 summarize the evolution of the system as a function of $V_\mathrm{BG}$ and $B^*$. The overlaid black circles in Fig. 3(a) represent the expected $V_\mathrm{BG}$ positions of the CFs’ $i=1$ GR with the WC potential. The white ellipses mark the $V_\mathrm{BG}$ positions at which we observe maximum values of peaks A, B, and C. The observed positions of the maxima agree well with the black circles. We emphasize that the expected positions are calculated based on independent measurements of $p_\mathrm{T}$ and $p_\mathrm{B}$ [Fig. 2(e)]. Note that there is also a weakening of the $\nu=5/9$ FQHS in Fig. 3(a) at $V_\mathrm{BG}=-40$ V, consistent with what we would expect when the GR coincides with a (weak) FQHS. In Fig. 3(a), for $B^* >0$, we also observe a weakening of the $\nu=4/9$ FQHS minimum when $V_\mathrm{BG}=-40$ V. This is slightly larger than $V_\mathrm{BG}$ where we expect the GR maximum ($V_\mathrm{BG}=-43$ V). On the other hand, for $B^*>0$, when $V_\mathrm{BG}<-40$ V, we do not observe variations in the peak values of maxima separating FQHSs which we could associate with the GR features. This is puzzling, but consistent with previous results for CFs in single-layer electron systems in an antidot lattice where the GR maxima for $B^*>0$ were much weaker than for $B^*<0$ \cite{Kang.PRL.1993,Smet.PRB.1997}. The origin of this asymmetry remains unclear. Figure 3(b) captures the magnetoresistance measured at $T=275$~mK. Here the resistance changes associated with the GR of CF orbits with the WC lattice are absent. The contrast between Figs. 3(a) and (b) provides strong evidence that the resistance maxima observed at $T= 30$~mK in Fig. 3(a) [and Fig. 2(b)] are caused by the WC formed in the bottom-layer, and they disappear once the WC has melted at $T= 275$~mK. 

It is worth highlighting some important aspects of our bilayer hole system. In an \textit{electron} bilayer system, the GR features induced by the WC are observed at very low fillings ($0.002<\nu_\mathrm{WC}<0.02$) compared to the hole sample studied here ($0.07< \nu_\mathrm{WC}<0.1$). This is consistent with previous studies in single-layer systems, where the relatively higher filling for the formation of the WC in 2D holes was associated with the larger hole effective mass and the resulting Landau level mixing \cite{Santos.PRL.1992,Shayegan.Review.1997,Santos.PRB.1992}. The higher $\nu_\mathrm{WC}$ for holes allows us to observe the WC at relatively higher densities or, equivalently, at smaller WC periods. Indeed, in Fig. 2(b) data, the period of the WC at which we observe the maxima indicated by the three arrows ranges from 80 to 140 nm. Such a small period means that the $i=1$ GR in the CF layer would require a large $B^*$, thus placing the GR feature far from the half-filling of the CF layer and well into its strong FQHS regime. Note that, because a very high-quality, modulation-doped, GaAs 2D carrier system containing CFs is typically buried deep under the surface, it would be extremely challenging to use structures (e.g., gates or antidots) patterned on the surface of the sample to impose periodic potentials with such small periods. Another major advantage of our bilayer system is that we can vary \textit{in-situ} the period of the periodic potential imposed on the CF layer via changing the WC layer density, and therefore tune the GR condition [Fig. 1(b)]. 

We also note that in our bilayer hole system, we do not observe higher-order ($i = 3, 7$) GR features seen in a bilayer electron system \cite{Deng.PRL.2016}. This might be because of the smaller mean-free-path of the CFs or the lack of long-range order in the WC layer in the hole system. We emphasize, however, that experiments on 2D electron systems with a 2D array of antidots whose positions deviate slightly from being periodic still show the GR feature associated with the $i=1$ orbit \cite{Gusev.JPCM.1994,Tsukagoshi.PRB.1995}. Evidently, the insensitivity of the $i=1$ GR to slight disorder in the periodic potential allows us to observe the primary GR maximum throughout a wide density range even when the period of the WC is not sharply defined.  

Returning to Fig. 2(b) data, the observation of GR of CFs near strong FQHSs implies that CFs possess a well-defined Fermi wavevector in this regime. This is at first sight surprising as one would expect that the CFs, which occupy a Fermi sea at $\nu=1/2$, should maintain a Fermi wavevector only at and near $\nu=1/2$. After all, in the CF picture, the FQHSs are considered to be the integer QHSs of the CFs \cite{Jain.PRL.1989,Jain.Book}, and the presence of strong FQHSs implies well-quantized CF Landau levels. On the other hand, theoretical studies imply that the concept of a Fermi wavevector for CFs appears to be valid even in the regime where strong FQHSs are expected \cite{Kamilla.PRB.1997,Barlam.PRL.2015}. 

Finally, our data appear to be consistent with the CF density being equal to $p_\mathrm{T}$, i.e., $k^*=\sqrt{4 \pi p_\mathrm{T}}$. If, as suggested in recent experiments \cite{Kamburov.PRL.2014}, we assume that the density of CFs for $B^*<0$ equals the $minority$ carriers in the lowest Landau level of the top-layer, then the expected positions for maxima in Fig. 2(b) are shifted to more negative values of $V_\mathrm{BG}$. Quantitatively, if we use values of $V_\mathrm{BG}$ at which we observe maxima in Fig. 2(b) and assume that the minority carriers are responsible for the CF GR, then we would obtain WC lattice constants that are about 15 to 20\% smaller than implied by our measured $p_\mathrm{B}$ [Fig. 2(e)]. We emphasize, however, while our $p_\mathrm{T}$ values are quite accurate, the values for $p_\mathrm{B}$ have reasonably large error bars because they are based on subtracting $p_\mathrm{T}$ from the extrapolated $p_\mathrm{Tot}$ in Fig. 2(e). Also, the range of $B^*$ showing GR in our study is much larger than those in \cite{Kamburov.PRL.2014} where CF GRs were observed very close to $\nu=1/2$. We conclude that the precise density of CFs that gives rise to GR away from half-filling remains an important open question, as also amplified in some very recent theoretical studies of the role of particle-hole symmetry in the CF formalism \cite{Barkeshli.PRB.2015,Barlam.PRL.2015,Son.PRX.2015,Wang.PRB.2016,Mulligan.PRB.2016,Cheung.PRB.2017,Wang.PRX.2017}.

\begin{acknowledgments}
We acknowledge support by the Department of Energy Basic Energy Sciences (Grant No. DE-FG02-00-ER45841) for measurements, and the National Science Foundation (Grants No. DMR 1305691, No. DMR 1709076, No. MRSEC DMR 1420541, and No. ECCS 1508925), and the Gordon and Betty Moore Foundation (Grant No. GBMF4420) for sample fabrication and characterization. M. S. also acknowledges support through a QuantEmX travel grant from The Institute for Complex Adaptive Matter and the Gordon and Betty Moore Foundation through Grant No. GBMF5305. We thank J. K. Jain and L. W. Engel for illuminating discussions.

\end{acknowledgments}


\begin{thebibliography}{99}
\bibitem{fnote1} Note that, based on measurements on single-layer samples, we estimate the bottom-layer resistance to be several hundred k$\Omega$ at high B, about 100 times larger than the top-layer resistance.
\bibitem{fnote2} For peak A, we also observe a small bump near $V_\mathrm{BG}=-20$~V [Fig. 2(b)]; we do not know its origin.
\bibitem{fnote3} The small step in $R_{1/2}$ near $V_\mathrm{BG}$ = -110 V is related to a slight change in the sample state during the measurements.
\end{thebibliography}
\end{document}